\documentclass[11pt]{article}
\usepackage{latexsym}  
\usepackage{amssymb}
\usepackage{graphicx}
\usepackage{amsmath}

\begin{document}
\begin{center}

{\Large\bf New Origin of a Bilinear Mass Matrix Form}
\vspace{2mm}

\vspace{3mm}
{\bf Naoyuki Haba and Yoshio Koide$^\dagger$}

{\it Department of Physics, Osaka University, 
1-1 Machikaneyama, Toyonaka, Osaka 560-0043, Japan} \\
{\it E-mail address: haba@het.phys.sci.osaka-u.ac.jp} \\

${}^\dagger${\it IHERP, Osaka University, 1-16 Machikaneyama, 
Toyonaka, Osaka 560-0043, Japan} \\
{\it E-mail address: koide@het.phys.sci.osaka-u.ac.jp}

\date{\today}
\end{center}

\begin{abstract}

The charged lepton mass formula can be explained 
 when the masses are propotinal to the squared 
 vacuum expectation values (VEVs) of 
 scalar fields. 
We introduce U(3) flavor symmetry and its nonet scalar field $\Phi$,
 whose VEV structure plays an essential role for generating the 
 fermion mass spectrum.  
We can naturally obtain bilinear form of the Yukawa coupling 
 $Y_{ij} \propto \sum_k \langle\Phi_{ik}\rangle
 \langle\Phi_{kj}\rangle$ without 
 the non-renormalizable interactions, when the flavor
 symmetry is broken only through 
 the Yukawa coupling and tadpole terms. 
We also speculate the possible
VEV structure of $\langle\Phi\rangle$.

\end{abstract}

\vspace{3mm}

The observed mass spectra of the quarks and leptons might provide 
 an important clue for the underlying theory. 
For the charged lepton sector, 
 we know the following empirical mass relation\cite{Koidemass,Koide90},
$$
m_e+m_\mu+m_\tau=\frac{2}{3}(\sqrt{m_e}+\sqrt{m_\mu}+\sqrt{m_\tau})^2 ,
\eqno(1)
$$
which can give a remarkable prediction $m_\tau = 1776.97$ MeV from 
the observed values of $m_e$ and $m_\mu$.
(The observed value is 
 $m_\tau^{obs}=1776.99^{+0.29}_{-0.26}$ MeV \cite{PDG06}.)
This mass relation seems to give remarkable hints for the origin
 of the mass spectrum.
In order to get the mass relation (1), 
 an interesting idea was proposed in Ref.\cite{Koide90}:
the mass spectrum originates not in the structure of the Yukawa 
 coupling constants $Y_{ij}$ but of the 
 vacuum expectation values (VEVs) $v_i$s of scalars $\phi_i$s as
$$
v_1^2+v_2^2+v_3^2 =\frac{2}{3}(v_1+v_2+v_3)^2 .
\eqno(2)
$$
Here we encounter following two questions.

(i) How can we obtain the VEV relation (2) naturally?

(ii) How to build a model in which 
 $m_{ei}$ has a bilinear form 
$$
 m_{ei} \propto v_i^2 ,
 \eqno(3)
$$
\hspace*{1.5cm}naturally?

\noindent
The first question seems to be related to a permutation symmetry
 of S$_3$\cite{S3}
 or higher symmetries which contain S$_3$.
The second question can be solved by the seesaw-type mass
 generation mechanism for the charged fermions \cite{UnivSeesaw}.
However, in the seesaw-type model, 
 we must identify the scalar $\phi$ as
 the three Higgs doublets with ${\mathcal O}(10^2)$ GeV VEVs,
 which may induce 
 the unwanted 
 large flavor changing neutral current (FCNC) \cite{FCNC}.  
On the other hand, 
 $\phi$s are not Higgs doublets in 
 the Froggatt-Nielsen-type model\cite{Froggatt} 
 so that 
 the FCNC problem might be avoided. 
However, it should be emphasized that 
 the bilinear form is just an assumption 
 in the Froggatt-Nielsen-type model.

The purpose of this paper is to propose a new mechanism 
 which induces the bilinear form $ m_{ei} \propto v_i^2$ 
 in the framework of a SUSY scenario.
The SUSY model which leads to the VEV relation (2) and 
 the bilinear form  
 has been firstly proposed by Ma \cite{Ma-PLB07}, where 
 four Higgs fields 
 $(\eta_i, \xi_i, \zeta_i, \psi_i)$ were introduced\footnote{
$\eta_i$, $\xi_i$, $\zeta_i$, and $\psi_i$ are 
 SU(2)$_L$-doublet Higgs fields,
 and 
 $\eta_i$ has the Yukawa interaction $f \eta_i L_i E_i$.}.  
The bilinear structure $m_{ei} \propto v_i^2$ has been realized 
 via $m_{ei} \propto \langle \eta_i^0\rangle
\propto \langle\zeta_i^0\rangle \langle\sigma_i\rangle
\propto \langle\sigma_i\rangle^2$, where $\langle\sigma_i\rangle$
satisfies the VEV relation (2).
This model is well organized but there are too many
 Higgs doublets. 
In this paper, 
 we will try to construct a new model which naturally induces the bilinear
 form of $Y_ij \propto \sum_k \langle\Phi_{ik}
 \rangle \langle\Phi_{kj}\rangle$ in 
 the different way from
 Refs.\cite{Ma-PLB07} and \cite{Koide0705}. 
We will introduce only one SU(2)$_L$-singlet
superfield $\Phi$ which plays a role of giving the VEV relation
(2) in addition to the conventional set of Higgs doubles, 
 $H_d$ and $H_u$, which give the masses of the charged
 leptons (and also the down-quarks) and the neutrinos (and
 also the up-quarks), respectively.

Under an flavor symmetry, 
 the leptons $L_i$ and $E_i$ are transformed as
$$
L=U_X L', \ \ \ E=U_X E' ,
\eqno(4)
$$
where $L_i$ and $E_i$ are the left-handed SU(2)$_L$ doublets
and the SU(2)$_L$ singlets, respectively.
We do not specify whether the transformation $U_X$ is
 continuous or discrete. 
In the conventional model, the Yukawa interaction of the
charged lepton sector is given by
$$
W_Y = \sum_{i,j} Y_{ij} L_i H_d E_j = {\rm Tr}[Y (EH_d L)].
\eqno(5)
$$
The Yukawa coupling constants $Y_{ij}$
 are strictly constrained by the symmetry under $U_X$, or 
 the symmetry is badly broken by the Yukawa interaction (5).
We would like to consider the 
 structure-less Yukawa coupling,      
 and the mass spectrum originates
 not in the Yukawa coupling constants $Y$ but in the VEV of
 scalars. 
In order for the Yukawa interactions to be invariant under
 the transformation $U_X$,
 we introduce 
 a nonet scalar $\Phi$ which transforms as
$$
\Phi=U_X \Phi' U_X^\dagger .
\eqno(6)
$$

When the flavor symmetry is U(3), the scalar $\Phi$ is regarded as 
 a nonet. 
A prototype model with a U(3) nonet scalar is found 
 in Ref.\cite{Koide90}, and  
 a more realistic U(3) nonet model is 
 proposed in Ref.\cite{Koide0705}.
The general form of $W_\Phi$ is given by
$$
W_\Phi = m_1 {\rm Tr}[\Phi\Phi] + m_2 ({\rm Tr}[\Phi])^2
+\lambda_1 {\rm Tr}[\Phi\Phi\Phi]
+\lambda_2 {\rm Tr}[\Phi\Phi] {\rm Tr}[\Phi] +
\lambda_3({\rm Tr}[\Phi])^3.
\eqno(7)
$$
A suitable choice of the parameters might give 
 non-zero magnitude of 
 $\langle\Phi\rangle$, and 
 an effective
 Yukawa interaction
 can be induced from 
$$
y \frac{1}{M} {\rm Tr}[\Phi (EH_d L)],
\eqno(8)
$$
which is invariant under the transformation of $U_X$.
This is a Froggatt-Nielsen-type model 
 proposed in Ref.\cite{Koide0705}.
The interaction (8) is a higher dimensional term
 which is accompanied with an energy scale
 $M$ of the effective theory, and 
 the bilinear form is not derived.
We will seek for another mechanism which can
 give $m_{ei}\propto v_i^2$ through the renormalizable 
 interactions.

Conventional models have considered 
 exact unbroken flavor symmetries at the beginning, which are 
 spontaneously broken later.
In this paper we take a different setup where 
 the superpotential $W$ has explicit ($U_X$) symmetry breaking terms, 
 which are common in Yukawa 
 interaction (4) and a tadpole terms 
 ${\rm Tr}[Y\Phi]$ as 
$$
W= W_\Phi -\mu^2 {\rm Tr}[Y \Phi] + W_Y .
\eqno(9)
$$
This shows 
$$
\frac{\partial W}{\partial \Phi} = 0 =\frac{\partial W_\Phi}{\partial \Phi}
-\mu^2 Y 
= 3 \lambda_1 \Phi \Phi + f_1 (\Phi) \Phi + f_0(\Phi) {\bf 1}-\mu^2 Y ,
\eqno(10)
$$
where
$$
f_1(\Phi)= 2(m_1 +\lambda_2 {\rm Tr}[\Phi] ) ,
\eqno(11)
$$
$$
f_0(\Phi)=2 m_2 {\rm Tr}[\Phi] +
\lambda_2 {\rm Tr}[\Phi\Phi] +3\lambda_3 ({\rm Tr}[\Phi])^2,
\eqno(12)
$$
and ${\bf 1}$ is a $3\times 3$ unit matrix.
Now we put an ansatz that our vacuum is given 
by the solution of Eq.(10) as  
$$
3 \lambda_1 \Phi \Phi -\mu^2 Y =0 ,
\eqno(13)
$$
and
$$
f_1 (\Phi) \Phi + f_0(\Phi) {\bf 1} =0.
\eqno(14)
$$
The requirement (13) realizes the bilinear 
 relation of our goal as 
$$
Y_{ij} = \frac{3\lambda_1}{\mu^2} \sum_k \langle\Phi_{ik}\rangle 
\langle\Phi_{kj}\rangle .
\eqno(15)
$$
For the existence of non-zero and non-degenerate eigenvalues of 
 $v_i$,
 Eq.(14) requires 
 $f_1=0$ and $f_0=0$, i.e.
$$
{\rm Tr}[\Phi] = -\frac{m_1}{\lambda_2} ,
\eqno(16)
$$
and
$$
2 m_2 {\rm Tr}[\Phi] +
\lambda_2 {\rm Tr}[\Phi\Phi] +3\lambda_3 ({\rm Tr}[\Phi])^2 =0.
\eqno(17)
$$
Since the value of $\langle\Phi\rangle$ is of the order of
$m_1/\lambda_2$, the Yukawa coupling constant $Y$ is of the
order of $m^2_1/\mu^2$.

Now let us consider 
 how to obtain the VEV relation (2). 
When we denote the nonet $\Phi$ in terms of the octet 
$\Phi^{(8)}=\Phi -\frac{1}{3}{\rm Tr}[\Phi]$ and the singlet
$\Phi^{(1)}=\frac{1}{3}{\rm Tr}[\Phi]{\bf 1}$\footnote{Notice
 that ${\rm Tr}[\Phi^{(8)}]=0$.}, 
 the term 
 ${\rm Tr}[\Phi\Phi\Phi]$ is devided into 
 the following two parts, 
$$
{\rm Tr}[\Phi\Phi\Phi]= {\rm Tr}[\Phi^{(8)}\Phi^{(8)}\Phi^{(8)}]
+{\rm Tr}[3\Phi^{(1)}\Phi^{(8)}\Phi^{(8)}+\Phi^{(1)}\Phi^{(1)}\Phi^{(1)}],
\eqno(18)
$$
$$
{\rm Tr}[\Phi^{(8)}\Phi^{(8)}\Phi^{(8)}]
={\rm Tr}[\Phi\Phi\Phi]-{\rm Tr}[\Phi] \left({\rm Tr}[\Phi\Phi]-
\frac{2}{9}({\rm Tr}[\Phi])^2\right) ,
\eqno(19)
$$
$$
{\rm Tr}[3\Phi^{(1)}\Phi^{(8)}\Phi^{(8)}+\Phi^{(1)}\Phi^{(1)}\Phi^{(1)}]
={\rm Tr}[\Phi] \left({\rm Tr}[\Phi\Phi]-
\frac{2}{9}({\rm Tr}[\Phi])^2\right) .
\eqno(20)
$$
As shown in Ref.\cite{Koide0705}, 
by imposing the Z$_2$ invariance 
 (Z$_2$ parities $+1$ and $-1$ are assigned to
 the fields $\Phi^{(1)}$ and $\Phi^{(8)}$, respectively),
the component
${\rm Tr}[\Phi^{(8)}\Phi^{(8)}\Phi^{(8)}]$ with the negative parity 
is dropped
from the term ${\rm Tr}[\Phi\Phi\Phi]$ 
 which 
 induces the VEV relation (2). 
Unfortunately,
 we cannot apply this Z$_2$ symmetry to our model
 because it derives 
 $\lambda_1=0$. 
So we just assume 
 that the cubic term is given by 
 Eq.(19) as\footnote{The form (21) is only a phenomenological assumption.
} 
$$
W_\Phi = m {\rm Tr}[\Phi\Phi] 
+ \lambda {\rm Tr}[\Phi^{(8)}\Phi^{(8)}\Phi^{(8)}]
\eqno(21)
$$
in the present stage. 
Since the cubic term ${\rm Tr}[\Phi\Phi\Phi]$ in the expression (19)
can be canceled with the tadpole term $-\mu^2{\rm Tr}[Y\Phi]$, 
the remaining terms are essentially identical with the expression (20).
Then the assumption (21) gives
$$
m_1 = m, \ \ m_2=0, \ \ \ \lambda_1=\lambda, \ \  
\lambda_2=-\lambda, \ \  \lambda_3 =\frac{2}{9} \lambda ,
\eqno(22)
$$
which leads to the VEV relation
$$
{\rm Tr}[\Phi\Phi] = \frac{2}{3}({\rm Tr}[\Phi])^2 ,
\eqno(23)
$$
with Eq.(17).
The relation (23) is the VEV relation (2) on the
basis of $\langle\Phi_{ij}\rangle =\delta_{ij} v_i$.

Now, let us discuss the neutrino sector.
If 
 the same scalar $\Phi$
contributes to the neutrino sector, we cannot
explain the observed value \cite{solar,atm}
$$
R \equiv \frac{\Delta m^2_{solar}}{\Delta m^2_{atm}}
=\frac{(7.9^{+0.6}_{-0.5})\times 10^{-5} {\rm eV}^2}{
(2.74^{+0.44}_{-0.26})\times 10^{-3} {\rm eV}^2}
=(2.9\pm 0.5) \times 10^{-2} ,
\eqno(24)
$$
because this gives too small value of
$R \simeq (m_\mu/m_\tau)^2 =3.4 \times 10^{-3}$ 
for Dirac neutrinos $m_i^{Dirac}\propto v_i^2 \propto m_{ei}$, 
and $R \simeq (m_\mu/m_\tau)^4 = 1.2 \times 10^{-5}$ 
for Majorana neutrinos with
$m_{\nu i} \propto (m_i^{Dirac})^2$.
So we should consider that the scalar $\Phi$ which
contributes to the neutrino sector is different from
the charged lepton sector (we will refer the former
as $\Phi_u$ and the latter as $\Phi_d$).
We would like to consider that the essential
structure of the superpotential $W(\Phi_u)$ is the same as
$W(\Phi_d)$ with the relation (23) for 
$\langle\Phi_u\rangle$.
Here, let us define a useful notation of dimensionless parameters 
$z_i$ which
is defined by $v_i= v z_i$, where $v=\sqrt{v_1^2+v_2^2+v_3^2}$.
Then, the values $z_i$s satisfy the relation
$z_1^2+z_2^2+x_3^2=1=(2/3)(z_1+z_2+z_3)^2$.
Remembering that three real solutions $x_i$s of a cubic equation
 $a x^3+b x^2 +c x +d=0$ are expressed by a form
 $x_i = \alpha + \beta \sin(\theta +(2/3)(i-1)\pi)$ ($i=1,2,3$), 
 the parameters $z_i$s can be expressed by
$$
\begin{array}{l}.
z_1 =\frac{1}{\sqrt6} -\frac{1}{\sqrt3} \sin\theta , \\
z_2 =\frac{1}{\sqrt6} -\frac{1}{\sqrt3} 
\sin\left(\theta+\frac{2}{3}\pi\right) , \\
z_3 =\frac{1}{\sqrt6} -\frac{1}{\sqrt3} 
\sin\left(\theta+\frac{4}{3}\pi\right) ,
\end{array}
\eqno(25)
$$
since $v_i$ are eigenvalues of the $3\times 3$ matrix 
 of 
$\langle\Phi\rangle$\footnote{
The factor $1/\sqrt6$ is coming from 
 the normalization of 
 $(z_1+z_2+z_3)^2=3/2$.}. 
Thus the ratio of (24) is written as 
$$
R_n =\frac{z_2^n -z_1^n}{z_3^n-z_2^n} .
\eqno(26)
$$
If neutrino masses are Dirac type 
without a seesaw mechanism, 
 the
 observed ratio (24) is given by Eq.(26) with $n=4$.
On the other hand, if neutrino masses
are Majorana type which are generated by a seesaw mechanism with
$M_R \propto {\bf 1}$, the ratio is given by Eq.(26) with $n=8$.
They suggest 
$$
\theta_\nu = 57.0^\circ \pm 1.4^\circ ,
\eqno(27)
$$
for $R_4=0.029\pm 0.005$ and
$$
\theta_\nu = 72.5^\circ \pm 0.8^\circ ,
\eqno(28)
$$
for $R_8=0.029\pm 0.005$\footnote{Here 
we chose the case $z_1^2 <z_2^2 \ll z_3^2$. 
Since we have not fixed the neutrino mixing matrix
 so far, we can also choose another solutions of $\theta_\nu$ 
 by the replacement of $\theta \rightarrow 60^\circ -\theta$ which
 corresponds to the case $z_2^2 <z_1^2 \ll z_3^2$.}. 
As for the charged lepton sector, 
 the observed charged lepton masses $(m_e, m_\mu, m_\tau)$ suggest
$$
\theta_e=42.7324^\circ ,
\eqno(29)
$$
which give $z_1=0.016473$, $z_2=0.236869$ and
$z_3=0.971402$.
It is interesting that the value (28) satisfies 
$\theta_\nu -\theta_e \simeq 30^\circ$.

So far, we have not discussed the neutrino mixing.
Notice that the results (16) -- (17) (and also (23)) are
 satisfied 
 independently of the flavor basis.
The Yukawa coupling constants $Y_\nu$ and
$Y_e$ are related to the VEV relations $\langle\Phi_f\rangle$ 
($f=u,d$) as 
$$
Y_\nu = \frac{3\lambda_u}{\mu_u^2} \langle \Phi_u\rangle^2 ,
\ \ \ 
Y_e = \frac{3\lambda_d}{\mu_d^2} \langle \Phi_d\rangle^2 .
\eqno(30)
$$
So if we fix the flavor basis of $L_i$, 
the basis of $(Y_\nu)_{ij}$ and $(Y_e)_{ij}$ are also fixed. 
For an example, if we choose the flavor basis in which $Y_e$
 is diagonal ($\langle \Phi_d\rangle$ is diagonal),
 the matrix $Y_\nu$ ($\langle \Phi_u\rangle$)
 is not diagonal on this basis in general.
So far 
 we can only know 
 the eigenvalues of $\langle\Phi_f\rangle$ and 
 cannot know the explicit form of the matrix
 $\langle \Phi_u\rangle$.
In order to fix the flavor mismatch between $Y_\nu$ and $Y_e$, 
 we try to 
 introduce an additional term $\varepsilon {\rm Tr}[B_f\Phi_f]$ 
 in the
 superpotential from the practical point of view as 
$$
W_f= W_{\Phi_f} -\mu^2_f {\rm Tr}[Y_f \Phi_f] + W_{Y_f} + 
\varepsilon {\rm Tr}[B_f\Phi_f],
\eqno(31)
$$
where $B_f$ are not fields 
 but numerical matrices. 
We assume that the basis where the VEV matrix
$\langle \Phi_f\rangle$ becomes diagonal is fixed by the
condition
$$
{\rm Tr}[B_f \Phi_f] = {\rm Tr}[U^\dagger_f B_f U_f 
\widetilde{\Phi}_f]=0 ,
\eqno(32)
$$
where
$$ 
\widetilde{\Phi}_f \equiv {\rm diag}(v_{f1}, v_{f2}, v_{f3}) 
= U^\dagger_f \Phi_f U_f .
\eqno(33)
$$
Since the matrices $B_f$ have been introduced 
only for the purpose to fix the flavor basis for 
the concerned Yukawa interaction, 
we can take $\varepsilon \rightarrow 0$ in the final results.
For an example, let us examine the case of \cite{Koide0705}, where 
the flavor symmetry is U(3)
 and it breaks to S$_4$.
The nonet scalar $\Phi_d$ is expected to be broken to 
${\bf 1}+{\bf 2} +{\bf 3}+{\bf 3}'$ of S$_4$ and 
the components of ${\bf 1}+{\bf 2}$ generate the charged
 lepton masses.
This splitting between ${\bf 1}+{\bf 2}$ and ${\bf 3}+{\bf 3}'$ 
is realized by a matrix
$$
B_e=\left(
\begin{array}{ccc}
0 & 1 & 1 \\
1 & 0 & 1 \\
1 & 1 & 0
\end{array} \right) ,
\eqno(34)
$$
because 
the components $\Phi_{ij}$ ($i\neq j$) denote ${\bf 3}+{\bf 3}'$ 
 of S$_4$ in the nonet expression of $\Phi$, 
 and the components
$\Phi_{11} = \frac{1}{\sqrt{3}} \Phi_\sigma + 
\frac{2}{\sqrt{6}} \Phi_\eta$, 
$\Phi_{22} = \frac{1}{\sqrt{3}} \Phi_\sigma - 
\frac{1}{\sqrt{6}} \Phi_\eta - \frac{1}{\sqrt{2}} \Phi_\pi$ and
$\Phi_{33} = \frac{1}{\sqrt{3}} \Phi_\sigma - 
\frac{1}{\sqrt{6}} \Phi_\eta + \frac{1}{\sqrt{2}} \Phi_\pi$ 
denote a singlet $\Phi_\sigma$ and a doublet $(\Phi_\pi, \Phi_\eta)$ of S$_4$.
In this case, 
 the trace ${\rm Tr}[B_e \tilde{\Phi_d}]$ 
is obviously zero with $U_e ={\bf 1}$.
As for the neutrino sector, the splitting
 between the doublet of S$_4$ is crucial 
 so we
 take
$$
B_\nu=\left(
\begin{array}{ccc}
0 & 0 & 0 \\
0 & -1 & 0 \\
0 & 0 & +1
\end{array} \right) ,
\eqno(35)
$$
which suggests $\phi_\pi$ is the component of the doublet
$(\phi_\pi, \phi_\eta)$ of S$_4$ as in Eq.(33).
The matrix $B_\nu$ is rotated by 
$$
U_\nu= U_{TB} \equiv \left(
\begin{array}{ccc}
\frac{2}{\sqrt6} & \frac{1}{\sqrt3} & 0 \\
-\frac{1}{\sqrt6} & \frac{1}{\sqrt3} & -\frac{1}{\sqrt2} \\
-\frac{1}{\sqrt6} & \frac{1}{\sqrt3} & \frac{1}{\sqrt2} 
\end{array} \right) ,
\eqno(36)
$$
as
$$
U_{TB}^\dagger B_\nu U_{TB}=
\frac{1}{\sqrt3} \left(
\begin{array}{ccc}
0 & 0 & -1 \\
0 & 0 & \sqrt2 \\
-1 & \sqrt2  & 0
\end{array} \right) ,
\eqno(37)
$$
where 
the flavor-basis-fixing term 
${\rm Tr}[B_\nu \Phi_u]={\rm Tr}[U^\dagger_\nu B_\nu U_\nu 
\widetilde{\Phi}_u]$
can be set to zero for $U_\nu=U_{TB}$.
It means that the Yukawa coupling constant 
$Y_\nu$ is given by 
$Y_\nu = (3\lambda/\mu^2) U_{TB} (\widetilde{\Phi}_u)^2 U_{TB}^\dagger$
on the basis where $Y_e$ is diagonal.
This suggests  the neutrino mixing matrix is given by 
``tri/bi-maximal mixing" 
$U_\nu =U_{TB}$ \cite{tribi}. 
Notice that this does not mean we have derived the tri/bi-maximal
mixing in our model, since 
 the mixing form is due to the ad hoc choice of 
 (35).
The ansatz (32) is only a trial, but 
 the introduction of a 
 flavor-basis-fixing term seems to be an interesting 
 candidate to complete our scenario.

In conclusion, we have examined 
 the idea that the fermion mass spectrum 
 originates not in the structure of the Yukawa coupling 
 but in the VEV structure. 
We have proposed a new mechanism 
 which gives a bilinear form of $m_i \propto v_i^2$ without 
 introducing higher dimensional interactions as in the 
 Froggatt-Nielsen model.  
We have applied this 
 mechanism to the charged lepton mass relation (2)
 at first. 
For the derivation of $Y\propto \langle\Phi\rangle^2$,
 it has been essential that the flavor symmetry of 
 the superpotential 
 $W_\Phi(\Phi)$ is broken only by
 the tadpole term $\mu^2{\rm Tr}[Y\Phi]$, 
where $\partial W/\partial \Phi =0$ has derived 
 $Y\propto\langle\Phi\rangle^2$.  
Notice that the bilinear form (15) is not a unique solution
 (vacuum), and  
 there are other solutions (vacuums) in the general 
 form of 
$$
Y =\frac{1}{\mu^2} \left\{ 3\lambda_1 \Phi\Phi
+ f_1(\Phi) \Phi + f_0(\Phi) {\bf 1}
\right\} .
\eqno(38)
$$
If we take the vacuum where the Yukawa 
coupling constant $Y$ is only proportional to 
$\langle \Phi \rangle$, i.e. $\mu^2 Y =2m_1\langle \Phi \rangle$,
we cannot obtain the non-degenerate and
non-zero eigenvalues of $\langle \Phi \rangle$.
The desirable eigenvalues (non-degenerate and non-zero
eigenvalues) exist in the vacuum of 
$\mu^2 Y =3\lambda_1\langle \Phi \rangle\langle \Phi \rangle$.
When we choose a solution of (13), 
we obtain $f_1=f_0=0$ as a byproduct in the present scenario.
Our purpose of this paper is not the derivation of the 
 formula of (2).
We have just assume the 
 form of (21),  
 which induces the VEV relation (2) through 
 the requirement of $f_0(\Phi)=0$. 

We have also applied the same mechanism to the neutrino sector.
We have shown 
 one attempt of generating the flavor mixings 
 by introducing the additional interaction.  
We will seek for more reasonable prescription of generating
 flavor mixings.
In this paper, we have not investigated the quark mass
spectra.
It is well known that the observed quark masses do not satisfy 
the relation (23) [(2)] (for example, see Table 1 in 
Ref.\cite{KoideJPG07}).
We will seek for a unified description including quark sectors
based on the bilinear mass matrix formulation.

\vspace{4mm}

\centerline{\large\bf Acknowledgments} 

One of the authors (NH) is supported by the Grant-in-Aid for
Scientific Research, Ministry of Education, Science and 
Culture, Japan (No.16540258 and No.17740146).
One of the authors (YK) is also supported by the Grant-in-Aid for
Scientific Research, Ministry of Education, Science and 
Culture, Japan (No.18540284).



\end{document}